\documentstyle[fleqn,a4,12pt]{article}

\newcounter{abc}

\newcounter{Rom}

\newcounter{rom}

\newcounter{nuljed}

\begin{document}

\vspace*{3ex}

\Large
\begin{center}

   \vspace*{2ex}
     {\bf  SUPERSYMMETRY AND QUBIT FIELD THEORY } \\

   \vspace*{2ex}
                   {\bf Jaroslav HRUB\'{Y}}

   {\bf Institute of Physics AV CR,Czech Republic }
   {\bf e-mail: hruby@gcucmp.cz  }

\end{center}

\vspace*{3ex}

\normalsize

\begin{abstract}
Supersymmetric extension of the Deutsch's quantum field theory is
presented and a new solution of quantum information paradox via
quantum-fields $ \bar{\psi} \psi $ condensate is presented.
\end{abstract}

\section{Introduction}

In recent time it is believed that quantum mechanics has the
potential to bring about a spectacular revolution in quantum
computing and quantum iformation theory \cite{1,2}.

Supersymmetry \cite{3} is subject of great interest among physicist and mathematicians
and play the fundamental role in particle physics and in gravity theories.Here
we show that it can play crucial role in quantum information theory, when energy
becomes sufficiently high and where Bekenstein's information bound playes role
\cite{4}.

In the 1970s theoretical physics developed a new fruitful concept
in supersymmetry, i. e. the concept of treating states and
superpartner states equally, where anticommuting $c$-numbers play the
important role.

The glue supersymmetric quantum mechanics and supersymmetry was first given by the
superposition Lagrangian in $(1+1)$ space-time dimensions \cite{5}
in~1977:
\begin{equation}
 L =
\frac{1}{2}[(\partial_{n}\phi)^{2}-V^{2}(\phi)+
        \bar{\psi}(i+V'(\phi))\psi]    \label{v1}
\end{equation}
where $\phi$ was a solitonic Bose field and $\psi$ was a Fermi
field and $V(\phi)$ was some nonlinear potential.

Substituting into (\ref{v1}) the following restriction to $(1+1)$
space-time dimension
\begin{eqnarray}
  & \phi\rightarrow x(t) & \partial_{n}\rightarrow\partial_{t} \\
  & \bar{\psi}\rightarrow\psi^{\top}\sigma_{2} &
    i\rightarrow i\partial_{t}\sigma_{2}
\end{eqnarray}
where \(\psi={\psi_{1}\choose\psi_{2}}\) with components being
interpreted as anticommuting $c$-numbers, $\sigma_{h}$ denotes
the Pauli matrices, then from (\ref{v1}) the Lagrangian of the
supersymmetric quantum mechanics is obtained.
\begin{equation}
 L_{SSQM}=\frac{1}{2}[(\partial_{t}x)^{2}-V^{2}(x)+
          \psi^{\top}(i\partial_{t}+\sigma_{2}V'(x))\psi]
                                                     \label{v2}
\end{equation}
and the corresponding Hamiltonian has the known form
\begin{equation}
 H_{SSQM}=\frac{1}{2}p^{2}+\frac{1}{2}V^{2}(x)
   +\frac{1}{2}i[\psi_{1},\psi_{2}]V'(x)  \label{v3}
\end{equation}
which was proposed as supersymmetric quantum mechanics
(SSQM) \cite {6}.
If we move from (0+1) dimension to (1+1) dimension we can define superfields.

Superfields provide an elegant and compact description of supersymmetry representations.
and are defined on superspace.We show it for the case (1+1) dimension and it can be extended
to higher dimensions as is usuall.

We shall define the superspace $E$ with  elements
\(z^{A}\in E,\ A=1,\ldots,8\), which are:
\begin{equation}
 (x^{\mu},\theta^{\alpha}) \ ; \qquad\mu=1,\ldots,4,\ \alpha=1,\ldots,4
          \ .                           \label{1}
\end{equation}

It can be viewed as a geometrical fiber space $E(V,W,P_{+}^{\dag})$, the basis $V$ is odd part
of the superspace and the fibre $W$ is the even part.
$P_{+}^{\dag}$ is the Poincar' group, acting on a fibre and E is the carthesian product between the basis and the fibre.

On the superspace we have the differential one-form:
\begin{eqnarray}
 \omega^{\mu} & = & \Omega^{\mu}+
      i\bar{\omega}^{\alpha}(\gamma^{\mu})_{\alpha\beta}\,\theta^{\beta}
                            \ , \label{2} \\
 \omega^{\alpha} & = & \alpha\,\theta^{\alpha} \ .     \label{3}
\end{eqnarray}

For the form \(\Omega^{\mu}=dx^{\mu}\) is valid the invariance
$\Omega_{\mu}\Omega^{\mu}$ under $P_{+}^{\dag}$.

From the fiber structure follows:
\begin{equation}
 \omega^{\mu} = \Omega^{\mu}+
      \bar{\omega}^{\alpha}\Gamma_{\alpha}^{\mu} \ , \label{2'}
\end{equation}
where $\Gamma_{\alpha}^{\mu}$ .

\[ \Gamma_{\alpha}^{\mu}=i(\gamma^{\mu})_{\alpha\beta}\,\theta^{\beta} \ . \]

It means $\varepsilon^{\alpha}$ in the anticommuting basis
\[ \theta^{\alpha} \rightarrow \theta^{\alpha}+\varepsilon^{\alpha} \]
makes change the $x^{\mu}$ on the
\[ x^{\mu} \rightarrow x^{\mu}+i\bar{\varepsilon}\gamma^{\mu}\theta \ . \]
 and it is the supersymmetry transformation, which preserve the invariance
 of the one-forms $\omega^{\mu}$ and $\omega^{\alpha}$.

We can see:
\begin{eqnarray*}
 \Omega^{\mu} & = & dx^{\mu}           \\
              & = & \left(\frac{\partial x^{\mu}}{\partial\theta^{\alpha}}
              \right) d\theta^{\alpha}         \\
              & = & \left(\frac{\partial x^{\mu}}{\partial\theta^{\alpha}}
              \right) \omega^{\alpha}  \ .
\end{eqnarray*}

Because the following is valid:
\( \frac{\partial x^{\mu}}{\partial \theta^{\alpha}}=
    \bar{\theta}'^{\beta}(\gamma^{\mu})_{\beta\alpha}\), \\
we obtain:
\(\Omega^{\mu}=\bar{\theta}'(\gamma^{\mu})\omega\).

If we know how to realize the Lorentz transformation in the basis
\[ \theta' \rightarrow S(\Lambda)\theta' \ , \qquad
  \omega \rightarrow S(\Lambda)\omega \ , \]
we obtain the Lorentz transformation on the fiber $x^{\mu}$.

We can see that the Lorentz transformation
\(x^{\mu}=\Lambda_{\nu}^{\mu}x^{\nu}\)
follows from the transformation $\Omega^{\mu}$:
\[ \Omega^{\mu} \rightarrow \Omega'^{\mu}
              \overline{S(\Lambda)\theta'}\gamma^{\mu}
      S(\Lambda)\omega=\theta'S^{-1}(\Lambda)\gamma^{\mu}
      S(\Lambda)\omega=\bar{\theta}'\Lambda_{\nu}^{\mu}\gamma^{\mu}
      \omega=\Lambda_{\nu}^{\mu}\omega^{\nu} \ . \]

In this way we obtained both the SUSY and the Lorentz
transformations on the superspace $E(V,W,P_{+}^{\dag})$.

Now we can ask what is the covariant derivative on the superspace?
From the differential geometry we know, that the covariant derivative is

\begin{equation}
 d Y^{J}+Y_{\mu}^{J}\omega^{\mu} = \bar{\omega}^{\alpha}Y_{\alpha}^{J} \ .
                      \label{4}
\end{equation}

If the object $Y^{J}$ transforms as the scalar under the supersymmetry transformation
, we cal it the ``superfield'' i.e. the local field
\(x^{\mu}\in W\) and \(\theta^{\alpha}\in V\).The index $J$
denotes the transformation under the Lorentz group; we suppose the scalar superfield
$\Phi(x,\theta)$.

we also get:
\begin{equation}
 \nabla \Phi(x,\theta) = d\Phi(x,\theta)+\Phi_{\mu}(x,\theta)+\Omega^{\mu}
        \ .              \label{4'}
\end{equation}

and

\[ \Omega^{\mu} = \omega^{\mu}-\bar{\omega}^{\alpha}\Gamma_{\alpha}^{\mu} \]
a dosad¡me-li tento v˜raz do rovnice (\ref{4'}), dostaneme:
\begin{equation}
 \nabla \Phi(x,\theta) = d\Phi(x,\theta)+\Phi_{\mu}(x,\theta)
 (\omega^{\mu}-\bar{\omega}^{\alpha}\Gamma_{\alpha}^{\mu})
        \ .              \label{5}
\end{equation}

So we obtain:
\[ \Phi_{\mu}(x,\theta)\omega^{\mu}=
    \bar{\omega}^{\alpha}\Phi_{\alpha}(x,\theta)-d\Phi(x,\theta) \]
and
\begin{eqnarray}
  \nabla \Phi(x,\theta) & = & \bar{\omega}^{\alpha}\Phi_{\alpha}(x,\theta)
      -\Phi_{\mu}(x,\theta)\,\bar{\omega}^{\alpha}\Gamma_{\alpha}^{\mu}
      \nonumber     \\
        & = & \bar{\omega}^{\alpha}(\Phi_{\alpha}(x,\theta)
      -\Phi_{\mu}(x,\theta)\,\Gamma_{\alpha}^{\mu})      \label{6}     \\
        & = & \bar{\omega}^{\alpha}\Phi_{i^{\alpha}}(x,\theta) \ .
      \nonumber
\end{eqnarray}

So we get:
\[ \Phi_{i^{\alpha}}(x,\theta)=\partial_{\alpha}\Phi(x,\theta)
   - i(\gamma^{\mu}\theta)_{\alpha}\,\partial_{\mu}\Phi(x,\theta)=
   D_{\alpha}\Phi(x,\theta) \ , \]
where \( D_{\alpha}=\partial_{\alpha}
    -i(\gamma^{\mu}\theta)_{\alpha}\,\partial_{\mu}=\partial_{\alpha}
    -\Gamma_{\alpha}^{\mu}\,\partial_{\mu} \)
is the covariant derivative.

It is known [3] $\omega^{\mu}$ and
$\omega^{\alpha}$ are invariant under  the supertransformation:
\[ \delta\,x^{\mu}= \bar{\varepsilon}^{\alpha}\Gamma_{\alpha}^{\mu} \ ,
  \qquad \delta\,\theta^{\alpha}=\varepsilon^{\alpha}\]
and:
\[ [\delta,\delta']\,x^{\mu}= 2i\bar{\varepsilon}'\gamma^{\mu}\varepsilon \ ,
  \qquad [\delta,\delta']\,\theta^{\alpha}=0 \ , \]
where \(\varepsilon,\varepsilon'\in V\) a element
$2i\bar{\varepsilon}'\gamma^{\mu}\varepsilon$
is the real element from $W$, so the $[\delta,\delta']\,x^{\mu}$ is the infiniteimal translation.
In this way we obtain the superalgebra:
\begin{eqnarray*}
 \{Q_{\alpha},\bar{Q}_{\beta}\} & = & -2(\gamma^{\mu})_{\alpha\beta}\,P_{\mu}
     \ ,       \\
   \left[P_{\mu},Q_{\alpha}\right]  & = & 0 \ ,          \\
   \left[P_{\mu},P_{\nu}\right]  & = & 0 \ .
\end{eqnarray*}

where:
\[ P_{\mu}\sim -i\partial_{\mu} \ , \qquad [P_{\mu},x^{\nu}]=
     -i\delta_{\mu}^{\nu} \ . \]

and:
\[ Q_{\alpha}\sim \eta_{\alpha\beta}\partial^{\beta} \ ,
   \qquad \{\partial_{\alpha},\theta^{\beta}\}=\delta_{\alpha\beta} \ , \]
The symbol $\eta_{\alpha\beta}$ denotes the metric, and so \(\eta_{\alpha\beta}=iC_{\alpha\beta}\)
where $C$ is the charge conjugation.

We can expand the superfield in

\(\theta={\theta_{\alpha}\choose\bar{\theta}_{\dot{\alpha}}}\)

and

\(\theta_{\alpha_{1}}\cdots\theta_{\alpha_{N}}\)
so we get:
\begin{eqnarray}
 Q(x,\theta,\bar{\theta}) & = & q(x)+\theta\varphi(x)+
    \bar{\theta}\bar{\chi}(x)  \nonumber \\
    & + & \theta\theta\, m(x)+ \bar{\theta}\bar{\theta}\, n(x)
     + \theta\sigma^{\mu}\bar{\theta}\,v_{\mu}(x)          \label{7} \\
    & + & \theta\theta\bar{\theta}\,\bar{\lambda}(x)+
     \bar{\theta}\theta\theta\,\psi(x)+
     \theta\theta\bar{\theta}\bar{\theta}\,d(x)     \nonumber
\end{eqnarray}
and all higher powers of $\theta$, $\bar{\theta}$ vanish.

The coefficients in expansion are the ordinary Bose and Fermi fields.
The scalar chiral superfield is defined

\begin{equation}
  \bar{D}_{\dot{\alpha}}\Phi(x,\theta,\bar{\theta}) = 0 \ .     \label{8}
\end{equation}

and it means:
\[ \Phi(x,\theta,\bar{\theta}) =
          e^{\theta^{\alpha}\bar{\theta}^{\dot{\alpha}}
 \partial_{\alpha\dot{\alpha}}}\Phi(y,\theta) \ ,\]
where \( e^{\theta^{\alpha}\bar{\theta}^{\dot{\alpha}}
 \partial_{\alpha\dot{\alpha}}}=1+i\theta\sigma^{\mu}\bar{\theta}
 \partial_{\mu}+\frac{1}{4}\theta^{2}\bar{\theta}^{2}\Box \ , \)
and \( y^{\mu}=x^{\mu}+i\theta\sigma^{\mu}\bar{\theta}\).

This means
$\Phi(x,\theta,\bar{\theta})$ is the function of $\theta$,
but not the $\bar{\theta}$. The Bose variable is the function of
\(y^{\mu}=x^{\mu}+i\theta\sigma^{\mu}\bar{\theta}\),

because the following is valid:
\( \bar{D}_{\dot{\alpha}}y^{\mu}=0 \ , \qquad
     \bar{D}_{\dot{\alpha}}\theta=0\).

For the superfield in \(y,\ \theta\) is valid:
\begin{eqnarray}
 \Phi(y,\theta) & = & A(y)+\sqrt{2}\theta\psi(y)+ \theta\theta\,F(y) \ ,
       \nonumber \\
    & = & A(x) + i\theta\sigma^{\mu}\bar{\theta}\partial_{\mu}A(x)
    +\frac{1}{4}\theta\theta\bar{\theta}\bar{\theta}\,\Box A(x)
                                                           \label{9} \\
    & + & \sqrt{2}\theta\psi(x)-\frac{i}{\sqrt{2}}\theta\theta
    \partial_{\mu}\psi(x)\sigma^{\mu}\bar{\theta}+ \theta\theta\,F(x)
    \ ,           \nonumber
\end{eqnarray}
IT can be obtained using derivation
 \(\theta,\ \theta\) in the \(\theta=\bar{\theta}=0\):
\begin{eqnarray*}
 \Phi\mid_{\theta=0} & = & A(x) \ ,     \\
 D_{\alpha}\Phi\mid_{\theta=0} & = & \sqrt{2}\psi_{\alpha}(x) \ , \\
 D_{\alpha}D^{\alpha}\Phi\mid_{\theta=0} & = & F(x) \ .
\end{eqnarray*}

We can also see:
\begin{eqnarray*}
 D_{\alpha} & = & \frac{\partial}{\partial\theta^{\alpha}}+
        2i\bar{\theta}^{\dot{\alpha}} \sigma_{\alpha\dot{\alpha}}^{\mu}
        \frac{\partial}{\partial y^{\mu}}  \ ,      \\
 \bar{D}_{\dot{\alpha}} & = &
       -\frac{\partial}{\partial\bar{\theta}^{\dot{\alpha}}} \ .
\end{eqnarray*}

The same way we can it obtain for the anticiral field:
\begin{equation}
 D_{\alpha}\bar{\Phi}(x,\theta,\bar{\theta})=0 \ .     \label{10}
\end{equation}

It can be see that chiral superfields are connected with the complex superfield
.
If we denote $R^{4|4}$ real superspace with the 4 real Bose variable
and 4 real Fermiho variable (like the
Majoran spinor), the cirality conditions represent
in $R^{4|4}$ superspace the extension of the real variable $x^{\mu}$
by the imaginary part $\pm i\theta\sigma^{\mu}\bar{\theta}$.

We move from the Bose variable $x^{\mu}$
to the \(y\equiv x_{L}^{\mu}=x^{\mu}+i\theta\sigma^{\mu}\bar{\theta}=
 (x^{\mu}-i\theta\sigma^{\mu}\bar{\theta})^{*}=(x_{R}^{\mu})^{*}\).

We can define the complex superspace $C^{4|2}$, which is parametrized by the
complex 4-vector $x_{L}^{\mu}\ (x_{R}^{\mu})$
and only one two component spinor
$\theta^{\alpha}\ (\bar{\theta}^{\dot{\alpha}})$,
as $x_{L}^{\mu},\theta^{\alpha}$,
then we can see $R^{4|4}$, as real hypersurface
\(Im\,x_{L}^{\mu}=\theta\sigma^{\mu}\bar{\theta}\)
in complex space $C^{4|2}$ (with the 2x4 + 2x2 real variables).

In the complex superspace the supersymmetry transformation has the form
\[ x_{L}^{\mu'}=x^{\mu}+2i\,\theta\sigma^{\mu}\bar{\varepsilon} \ , \qquad
 \theta^{\alpha'}=\theta^{\alpha}=\varepsilon^{\alpha} \ .      \]

The advantage is that here the role playes only
 $\theta$, but not $\bar{\theta}$, as in the $R^{4|4}$.

The superanalycity and antisuperanalycity will play important role in extended superszmetries.

The construction of the supersymmetry Lagrangians as is usual \cite{3}:
\begin{eqnarray*}
 \delta\int d^{4}x\,L & = & \int d^{4}x \left(\bar{\varepsilon}
 \frac{\partial}{\partial\bar{\theta}}+\frac{i}{2}
 (\bar{\varepsilon}\gamma_{\mu}\theta)\frac{\partial}{\partial x_{\mu}}L
 \right)         \\
 & = & \bar{\varepsilon} \frac{\partial}{\partial\bar{\theta}}\int d^{4}x\,L
 + \mbox{``surface term ``}=0 \ .
\end{eqnarray*}

and for ordinary fields  we can obtain \cite{3} :
\begin{eqnarray*}
 L & = & i\partial_{\mu}\bar{\psi}\bar{\sigma}^{\mu}\psi+A_{i}^{*}\Box A_{i}+
  F_{i}^{*} F_{i}            \\
 & + & \left[m_{ij}(A_{i}F_{j}-\frac{1}{2}\psi_{i}\psi_{j})\right.   \\
 & + &  g_{ijk}(A_{i}A_{j}F_{k}-\psi_{i}\psi_{j}A_{k})
     +\lambda_{i}F_{i}+ {\rm h.c.}  \biggr] \ .
\end{eqnarray*}

The auxiliary fields may be eliminated through their Euler equations:
\begin{eqnarray*}
 \frac{\partial L}{\partial F_{k}^{*}} & = &
   F_{k}+\lambda_{k}^{*}+m_{ik}A_{i}^{*}+g_{ijk}^{*}A_{i}^{*}A_{j}^{*} = 0
   \ ,           \\
 \frac{\partial L}{\partial F_{k}} & = &
   F_{k}^{*}+\lambda_{k}+m_{ik}A_{i}+g_{ijk}A_{i}A_{j} = 0
   \ ,
\end{eqnarray*}
Then $L$ can be express as:
\begin{eqnarray*}
 L & = & i\partial_{\mu}\bar{\psi}\bar{\sigma}^{\mu}\psi+A_{i}^{*}\Box A_{i}+
  -\frac{1}{2} m_{ik}\psi_{i}\psi_{k}
     -\frac{1}{2} m_{ik}^{*}\bar{\psi_{i}}\bar{\psi_{k}}          \\
 & - &  g_{ijk}\psi_{i}\psi_{j}A_{k}
        -g_{ijk}^{*} m_{ik}^{*}\bar{\psi_{i}}\bar{\psi_{j}}A_{l}^{*}
          - U(A_{i},A_{i}^{*}) \ ,
\end{eqnarray*}

where \(\ U(A_{i},A_{i}^{*})=F_{i}^{*}F_{i}\).

In this paper in Sect.2 we repeat some basic ideas from quantum information theory.

In Sect.3 we define a form for the antiqubits on the quantum mechanical level,
but the full sense it has in the anticommuting qubit field level.

In Sect.4,5 we show the role of qubits for Feymann diagrams and error correction
codes in quantum computing.Alsowe present a new interesting idea of the
application the superpartner potentials and partner--superpartner
energy levels as two level system for quantum computing and we
show an analog between supersymmetric square root and square root of
not.

In Sect.6 we show the possibility of the existence the information qubit quarks.

In sect.7 we present supersymmetric extension of the qubit field theory which has been presented
by D.Deutsch \cite{8}.

\section{Some ideas of quantum information theory}

       A quantum bit (qubit) is a quantum system with a two-dimensional
Hilbert space, capable of existing in a superposition of Boolean states
and of being entangled with the states of other qubits [1].

More precisely a qubit is the amount of the information which is contained
in a pure quantum state from the two-dimensional Hilbert space ${\cal H}_2$.

A general superposition state of the qubit is
\begin{equation}
 |\psi\rangle  = {\psi}_0 |0\rangle  + {\psi}_1 |1\rangle  ,
\label{1.1}
\end{equation}
where  ${\psi}_0$  and   ${\psi}_1$ are complex numbers, $|0\rangle$
and $|1\rangle$ are kets representing two Boolean states.
The superposition state has the propensity to be a $0$ or a $1$ and
${|{\psi}_0|}^2 + {|{\psi}_1|}^2 = 1$.

The eq.(1) can be written as
\begin{equation}
 |\psi\rangle  = {\psi}_0\left( \begin{array}{c} 1 \\ 0 \end{array} \right)
 + {\psi}_1 \left( \begin{array}{c} 0 \\ 1 \end{array} \right)  ,
\label{1.2}
\end{equation}

where we labeled $\left( \begin{array}{c} 1 \\ 0 \end{array} \right) $ and
$\left( \begin{array}{c} 0 \\ 1 \end{array} \right)$ two basis states zero
and one.

The Clifford algebra relations of the $2\times2$ Dirac matrices is
\begin{equation}
\left\{{\gamma}^\mu,{\gamma}^\nu \right\} = 2 {\eta}^{\mu\nu} ,
\label{1.3}
\end{equation}
where
\begin{equation}
 {\eta}^{\mu\nu}    =  \left( \begin{array}{cc}
                              -1 & 0 \\
                               0 & 1 \end{array} \right).
\label{1.4}
\end{equation}

We choose the representation
\begin{equation}
 {\gamma}^0 =i{\sigma}^2   =  \left( \begin{array}{cc}
                              0 & 1 \\
                             -1 & 0 \end{array} \right)
\label{1.5}
\end{equation}
and
\begin{equation}
 {\gamma}^1 ={\sigma}^3   =  \left( \begin{array}{cc}
                              1 & 0 \\
                              0 &-1  \end{array} \right) ,
\label{1.6}
\end{equation}
where $\sigma$ are Pauli matrices and ${\gamma}^5={\gamma}^0{\gamma}^1$ .

 The projectors have the form
\begin{equation}
P_0 = \frac{1+{\gamma}^1}{2} =  \left( \begin{array}{cc}
                              1 & 0 \\
                              0 & 0  \end{array} \right)
 \label{1.7}
\end{equation}
and
\begin{equation}
P_1 = \frac{1-{\gamma}^1}{2} =  \left( \begin{array}{cc}
                              0 & 0 \\
                              0 & 1  \end{array} \right)
 \label{1.7}
\end{equation}

These projectors project qubit on the basis states zero
and one:
\begin{equation}
P_0  |\psi\rangle  = {\psi}_0\left( \begin{array}{c} 1 \\ 0 \end{array} \right),
P_1  |\psi\rangle  = {\psi}_1\left( \begin{array}{c} 0 \\ 1 \end{array} \right)
\label{1.9}
\end{equation}
and represent the physical measurements - the transformations of qubits to the
classical bits.

In classical information theory the Shannon entropy is well defined :
\begin{equation}
S_{CL}(\Phi) =-\sum_{\phi} p(\phi) {\log}_2 p(\phi),
\label{1.10}
\end{equation}
where the variable $\Phi$ takes on value $\phi$ with probability  $p(\phi)$
and it is interpreted as the uncertainty about $\Phi$.

The quantum analog is the von Neumann entropy $S_{q}({\rho}_{\Psi})$ of a quantum
state $\Psi$ described by the density operator ${\rho}_{\Psi}$:
\begin{equation}
S_Q(\Psi) = - {Tr}_{\Psi}\left[ {\rho}_{\Psi} {\log}_2{\rho}_{\Psi} \right],
\label{1.11}
\end{equation}
where ${Tr}_{\Psi}$  denotes the trace over the degrees of freedom associated
with $\Psi$.
x
The von Neumann entropy has the information meaning, characterizing (asymptotically)
the minimum amount of quantum resources required to code an ensemble of quantum
states.

The density operator ${\rho}_{\psi}$ for the qubit state $ |\psi\rangle $ in (1)
is given:

\begin{equation}
\rho = |\psi\rangle \langle\psi| = {|\psi_0|}^2|0\rangle \langle0| +
\psi_0{\psi_1}^* |0\rangle \langle1| + {\psi_0}^*\psi_1 |1\rangle \langle0| +{|\psi_1|}^2|1\rangle \langle1|
\label{1.12}
\end{equation}
and corresponding density matrix is
\begin{equation}
{\rho}_{kl} = \left( \begin{array}{cc}
                              {|\psi_0|}^2 & \psi_0{\psi_1}^*  \\
                           {\psi_0}^*\psi_1 & {|\psi_1|}^2  \end{array} \right)
\label{1.13}
\end{equation}
and $k,l=0,1.$

The von Neumann entropy reduces to a Shannon entropy if ${\rho}_{\Psi}$ is a mixed
state composed of orthogonal quantum states.

In the interesting work [2], N.J.Cerf and C.Adami proposed the conditional
von Neumann entropy as follows:

for the combined system from two states $|i\rangle$  and $|j\rangle$
the von Neumann entropy
\begin{equation}
S_Q(|ji\rangle) = - Tr\left( {\rho}_{|ji\rangle} {\log}_2{\rho}_{|ji\rangle} \right),
\label{1.14}
\end{equation}
\begin{equation}
S_Q(|i\rangle) = - Tr\left( {\rho}_{|i\rangle} {\log}_2{\rho}_{|i\rangle} \right)
\label{1.15}
\end{equation}
and
\begin{equation}
{\rho}_{|j\rangle} = - {Tr}_{|i\rangle}\rho_{|ji\rangle}.
\label{1.16}
\end{equation}

The von Neumann conditional entropy has the form
\begin{equation}
S_Q(|j\rangle | |i\rangle) = - Tr\left( {\rho}_{|ji\rangle} {\log}_2{\rho}_{|j\rangle | |i\rangle} \right).
\label{1.17}
\end{equation}

   The appearence of negative values for the von Neumann conditional entropy
follows from (17), where the  conditional density matrix ${\rho}_{|j\rangle | |i\rangle}$
is based "conditional amplitude operator" [2]:
\begin{equation}
{\rho}_{|j\rangle | |i\rangle} = \exp \left(- \sigma_{|ji\rangle} \right),
\label{1.18}
\end{equation}
where
\begin{equation}
\sigma_{|ji\rangle} = I_{|j\rangle}\ln{\rho}_{|i\rangle} - \ln {\rho}_{|ji\rangle}
\label{1.19}
\end{equation}
and  $I_{|j\rangle}$ being the unit matrix.

Because the information cannot be negative the question arise what precisely
it means.

The quantum entropy of a given quantum state (1) in ${\cal H}_2$ is the difference
between the maximum of the information contained in (1) and information of the information vacuum.

For the qubit
\begin{eqnarray}
S_Q(|\psi\rangle | |v\rangle)=-Tr\left( {\rho}_{|\psi v\rangle} {\log}_2{\rho}_{|\psi\rangle | |v\rangle} \right) =\\
=S_Q(\psi|v)=S_Q(\psi v)- S_Q(v):=S_Q(\psi)=1,
\label{1.20}
\end{eqnarray}
where $|v\rangle$ is the information vacuum state $S_Q(v)=0$.

Now we can start to think about anti-quantum bits as the anti-quantum state in
the dual Hilbert space $\tilde{\cal H}_2$.

\section{Anti-quantum bits}
We shall define anti-quantum bits (antiqubits) by analogy with antiparticles,
as a quantum state from the dual
Hilbert space, capable of existing in a superposition of Boolean states
and of being entangled with the states of other qubits.

This definition agree formally with these previous ones in [2,3],
 where antiqubit is called as a quantum of negative information, which is
 equivalent to a qubit traveling backwards in time.

But here the information of antiqubit is positive only entropy $S_Q(|\bar{\psi\rangle} | |0\rangle)=S_Q(\bar{\psi}):=-1 $,
where $|\bar{\psi\rangle} $ denotes antiqubit state.

For the antiqubit
\begin{eqnarray}
S_Q(|\bar{\psi}\rangle | |v\rangle)=-Tr\left( {\rho}_{|\bar{\psi} v\rangle} {\log}_2{\rho}_{|\bar{\psi}\rangle | |v\rangle} \right)= \\
=S_Q(\bar{\psi}|v)=S_Q(\bar{\psi} v)- S_Q(v):=S_Q(\bar{\psi})=-1.
\label{1.21}
\end{eqnarray}

To obtain the form of $|\bar{\psi\rangle}$ we use the Dirac adjoint and old
ideas about anti-states from Particle Physics.

The antiparticles positrons was defined by Dirac as the energy holes so we
can define the antiqubits as the information holes. As the mass in the Dirac
eq. has the opposite sign here the entropy of qubit and antiqubit has the
opposite sign.

By analogy with the Dirac
adjoint for the two component spinor, we define antiqubit $\bar{\psi} $ as follows:
\begin{equation}
\bar{\psi}=\psi^+ {\gamma}^0 =\left({\psi_0}^*,{\psi_1}^* \right) \left( \begin{array}{cc}
                              0 & 1 \\
                             -1 & 0 \end{array} \right)
= \left({-\psi_1}^*,{\psi_0}^* \right) = {\psi_0}^* \langle 1| - {\psi_1}^* \langle 0| .
\label{2.1}
\end{equation}
The corresponding density matrix for the antiqubit $\bar{\psi}$ is
\begin{equation}
{{\rho}_{kl}}^` = \left( \begin{array}{cc}
                              {|\psi_0|}^2 & -\psi_0{\psi_1}^*  \\
                              -{\psi_0}^*\psi_1 & {|\psi_1|}^2  \end{array} \right)
\end{equation}
and $k,l=0,1.$

The information vacuum is $|v\rangle \in {\cal H}_2\otimes\tilde{\cal H}_2$.

The two members of a spatially separated Einstein-Podolsky-Rosen (EPR) pair are
maximally entangled qubits so called ebits [2,4].
EPR pair $e\bar{e}$ is created from the information vacuum, which is the pure
quantum state without information $S_Q(\psi_e\psi_{\bar{e}})=S_Q(\psi_e \bar{\psi_e})=S_Q(e\bar{e})=S_Q(|v\rangle)=0$.

As the EPR pairs $e\bar{e}$ after creation from the information
vacuum contain no readable information they represents the virtual information
elementary objects.
The conditional quantum entropy between $e$ and $\bar{e}$ is
 $S_Q(e|\bar{e})=S_Q(e\bar{e})-S_Q(\bar{e})=0-(-1)=1 $,

$S_Q(\bar{e}|e)=S_Q(\bar{e}e)-S_Q(e)=0-1=-1$.

Quantum information processes
can be described by quantum information diagrams similar to Feymann diagrams in
Particle Physics, where elementary objects interact.
The information flow is conserved in each vertex of these diagrams.
The law of the conservation of quantum information has no analog in classical
information

The elementary objects of the quantum information dynamics (QID) can be
classical bits $c$, qubits $q$, antiqubits $\bar{q}$, ebits $e$ and antiebits
 $\bar{e}$. But there could be some more elementary virtual quantum information
 objects "the information quarks".

Moreover these virtual elementary objects of the  QID can be combined into the
more complicated information objects in analogy with quarks in the bag models.
There is shown that quarks play no role for the energy of the bag [5] and so virtual
quantum information entangled quarks (equarks)
play no role for the information of the information bag.

\section{Feymann quantum information diagrams}

We show QID for the information processes teleportation  and  superdense
coding, which are connected via time reversal operation.

In quantum teleportation [see Fig.1] a qubit $\psi_q$ is transported with
perfect fidelity between two vertices $M$ and $U$ through the transmission
of two classical $2c$ bits and shared EPR pair $e\bar{e}$:

\setlength{\unitlength}{1mm}
\begin{picture}(40,50)
\put(12,20){$\psi_q$}
\put(33,29){$\psi_e$}
\put(33,9){$2c$}
\put(33,12){$\succ$}
\put(33,26){$\prec$}
\put(20,19){$\succ$}
\put(48,19){$\succ$}
\put(35,20){\circle{30}}
\put(18,20){\line(1,0){10}}
\put(28,20){\circle*{1}}
\put(42,20){\circle*{1}}
\put(52,20){\circle*{1}}
\put(18,20){\circle*{1}}
\put(42,20){\line(1,0){10}}
\put(30,19){$M$}
\put(38,19){$U$}
\put(55,20){$\psi_q$}
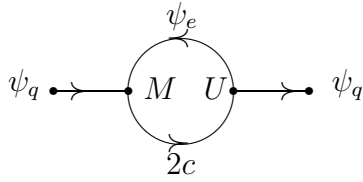
\put(1,1){Fig.1 Feymann quantum information diagram for teleportation.}
\end{picture}

Here $M$ (the measurement) and $U$ (the unitary
transformation) mean the two processes in QID and their description
is written elsewhere [6].
In vertex M qubit interacts with ebit and it gives four orthogonal maximally
entangled $|eq\rangle$ states, which are transported via 2 cbits.

In vertex U is the reconstruction of ${\psi}_q$ the  qubit from 2 c bits
via applying to antiebit one of four possible unitary transforms in the
one state Hilbert space ${\cal H}_2$.

The information flow is conserved in each vertex of the diagram on Fig.1.
For the vertex M the entropy conservation rule is
\begin{equation}
S_Q(q)+S_Q(e) = 1 + 1 = 2
\label{3.1}
\end{equation}
since qubit and ebit are initially independent.
At vertex U we have
\begin{equation}
S_Q(q)=S_Q(2c)+S_Q(\bar{e})=2-1=1=S_Q(qe\bar{e})=S_Q(qe) + S_Q(\bar{e}|qe).
\label{3.2}
\end{equation}
From eq.(2) we can see that outgoing $e$ from vertex U  is equivalent
to the incoming $\bar{e}$ to U.

In superdense coding [see Fig.2] we transport 2 cbits via 1 qubit.
The two cbits can be packed into one qubit via applying the unitary
transform what is equivalent to the interaction $2c$ with $\bar{e}$.
At vertex M the interaction between qubit and ebit gives recovering
of the 2 cbits.

\setlength{\unitlength}{1mm}
\begin{picture}(50,50)
\put(12,20){$2c$}
\put(33,29){$\bar{\psi_e}$}
\put(33,9){$\psi_q$}
\put(33,12){$\succ$}
\put(33,26){$\prec$}
\put(20,19){$\succ$}
\put(48,19){$\succ$}
\put(35,20){\circle{30}}
\put(18,20){\line(1,0){10}}
\put(28,20){\circle*{1}}
\put(42,20){\circle*{1}}
\put(52,20){\circle*{1}}
\put(18,20){\circle*{1}}
\put(42,20){\line(1,0){10}}
\put(30,19){$U$}
\put(37,19){$M$}
\put(55,20){$2c$}
\put(1,1){Fig.2 Feymann quantum information diagram for superdense coding.}
\end{picture}

The diagramm in Fig.2 is equivalent to the Feymann  quantum information diagram where ebit
$\psi_e$ is sent backwards in time ( it is equivalent to antiebit $\bar{\psi_e}$):

\begin{picture}(50,50)
\put(12,20){$2c$}
\put(33,29){$\psi_e$}
\put(33,9){$\psi_q$}
\put(33,12){$\succ$}
\put(33,26){$\succ$}
\put(20,19){$\succ$}
\put(48,19){$\succ$}
\put(35,20){\circle{30}}
\put(18,20){\line(1,0){10}}
\put(28,20){\circle*{1}}
\put(42,20){\circle*{1}}
\put(52,20){\circle*{1}}
\put(18,20){\circle*{1}}
\put(42,20){\line(1,0){10}}
\put(30,19){$U$}
\put(37,19){$M$}
\put(55,20){$2c$}
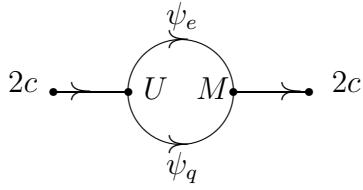
\put(1,1){Fig.3 Feymann quantum information diagram for superdense coding with $\psi_e$.}
\end{picture}

In this case at vertex U we have
\begin{equation}
S_Q(2c) + S_Q(\bar{e})= 2-1=1=S_Q(q|e)=S_Q(2c\bar{e}|e)=S_Q(2c)+S_Q(\bar{e}|e),
\label{3.3}
\end{equation}
since 2c and $\bar{e}$ are initially independent.
At M is
\begin{equation}
S_Q(2c)=S_Q(qe)=S_Q(q|e) +S_Q(e) = 1+1=2.
\label{3.3}
\end{equation}

The quantum information conservation law can have important consequences in
cosmology.

\section{Quantum error correction using antiqubits and  SSQM in
               quantum computing }

The main problem in quantum computing are error processes, such as decoherence
and spontaneous emission, on the state of a quantum memory register. For
example if a quantum memory register is described by an isolated qubit (1)
with the density matrix (13) at initial moment, the density matrix changes
over time:
 \begin{equation}
{\rho}_{kl} = \left( \begin{array}{cc}
                              {|\psi_0|}^2 &e^{-\frac{t}{\tau}}\psi_0{\psi_1}^*  \\
                          e^{-\frac{t}{\tau}} {\psi_0}^*\psi_1 & {|\psi_1|}^2  \end{array} \right)
\label{1.13}
\end{equation}
and $k,l=0,1.$

Here $\tau$, called the "decoherence time", sets the characteristic time-scale
of the decoherence process.

 For the antiqubit (21) we obtain

\begin{equation}
{{\rho}_{kl}}^` = \left( \begin{array}{cc}
                              {|\psi_0|}^2 & -e^{-\frac{t}{\tau}}\psi_0{\psi_1}^*  \\
                              -e^{-\frac{t}{\tau}}{\psi_0}^*\psi_1 & {|\psi_1|}^2  \end{array} \right)
\end{equation}
and $k,l=0,1.$

Quantum error-correction strategy can be based on the idea of mirror computers,
which are working in paraell regime with qubits and antiqubits.

It is known that quantum error-correcting code [7] works on the idea
that several physical qubits are used to encode one logical qubit.The actual
value of the logical qubit is stored in the correlations between the several
classical physical qubits so that even if there is disruption to a few
qubits in the codewords, there is still sufficient information in the
correlation when error occurr and then correct the qubit.

In mirror computation we can use the correlation between ebits and antiebits
to anihilate the error information.

We know that any error in a single bit can be described by the action
of linear combination of Pauli matrices:

\begin{equation}
{\sigma}_1 \left( \begin{array}{c} {\psi}_0 \\ {\psi}_1 \end{array} \right)=\left( \begin{array}{cc}
                               0 & 1 \\
                               1 & 0 \end{array} \right)\left( \begin{array}{c} {\psi}_0 \\ {\psi}_1 \end{array} \right)=\left( \begin{array}{c} {\psi}_1 \\ {\psi}_0 \end{array} \right),
\label{3.3}
\end{equation}
i.e. bit flip error,

\begin{equation}
{\sigma}_2 \left( \begin{array}{c} {\psi}_0 \\ {\psi}_1 \end{array} \right)=\left( \begin{array}{cc}
                               0 & -i \\
                               i& 0 \end{array} \right)\left( \begin{array}{c} {\psi}_0 \\ {\psi}_1 \end{array} \right)=i\left( \begin{array}{c} {\psi}_1 \\ {\psi}_0 \end{array} \right),
\label{3.3}
\end{equation}
i.e. phase shift plus bit flip error,

\begin{equation}
{\sigma}_3 \left( \begin{array}{c} {\psi}_0 \\ {\psi}_1 \end{array} \right)=\left( \begin{array}{cc}
                               1&0 \\
                               0&-1 \end{array} \right)\left( \begin{array}{c} {\psi}_0 \\ {\psi}_1 \end{array} \right)=\left( \begin{array}{c} {\psi}_0 \\-{\psi}_1 \end{array} \right),
\label{3.3}
\end{equation}
i.e. phase shift error.

For antiqubits we obtain the same in the dual space.

In general, an $N$-qubit can be in an arbitrary superposition of
all $2^{N}$ classical states:
\[ |\psi_{N}\rangle=\sum\alpha_{x}|x\rangle \ , \
     x\in\{|0x\rangle,|1\rangle\}^{N}\
           \mbox{and}\ \sum_{x}|\alpha_{x}|^{2}=1 \ . \]

It is known that two bit gates are universal
for quantum computation, which is likely to greatly simplify the
technology required to build quantum computers.

In our application the quantum interference can be given by the
superposition of $N$ two-level solitonic system.

To understand this we start with a simple idea of quantum algorithm
square root of not in the form from \cite{1} via partner--superpartner
case.

We begin with supersymmetric square root in SSQM.

SSQM is generated by supercharge operators $Q^{+}$ and
\(Q^{-}=(Q^{+})^{+}$ which together with the Hamiltonian $H=2H_{SSQM}$
of the system, where
\begin{eqnarray*}
 H_{SSQM} & = & \frac{1}{L}\left(
     \begin{array}{cc}
       -\frac{d^{2}}{d\,t^{2}}+v^{2}+v' & 0 \\
       0 & -\frac{d^{2}}{d\,t^{2}}+v^{2}-v'
     \end{array}  \right)    \\
 H & = & \left(
     \begin{array}{cc}
       H_{0} & 0 \\ 0 & H_{1}
     \end{array}  \right) =
   \left(
     \begin{array}{cc}
       A^{+}A^{-} & 0 \\ 0 & A^{-}A^{+}
     \end{array}  \right) =
     -\left( \frac{d^{2}}{dx^{2}}\right)+\sigma_{\beta}v'
\end{eqnarray*}

fulfil the superalgebra
\[(Q^{\pm})^{2}=0 \ , \ \ [H,Q^{-}]=[H,Q^{+}] \ ,
                              \ \ H=\{Q^{+},Q^{-}\}=Q^{2}\]
where
\[ Q^{-}=\left(
     \begin{array}{cc}
       0 & 0 \\  A^{-} & 0
     \end{array}  \right) \ , \
 Q^{+}=\left(
     \begin{array}{cc}
       0 & A^{+} \\  0 & 0
     \end{array}  \right)   \ , \  \  Q = Q^{+}+Q^{-} \]
and
\[ A^{\pm}=\pm\frac{d}{dx}+v(x) \ , \ \ v'=\frac{dv}{dx} \ .\]

Such Hamiltonians \(H_{0},H_{1}\) fulfil
\[H_{0}A^{+}=A^{+}H_{1} \ , \ \ A^{-}H_{1}=H_{0}A^{-}  \]

The coefficients $\alpha_{0}$ and $\alpha_{1}$ are called the
amplitudes of the $|0\rangle$ and $|1\rangle$, respectively.

The previous relations in lead to the double degeneracy of all positive
energy levels, of belonging to the ``0'' or ``1'' sectors specified
by the grading state operator $S=\sigma_{3}$, where
\[[S,H]=0 \ \mbox{and}\ \{S,Q\}=0 \ . \]

The $Q$ operator transforms eigenstates with \(S=+1\), i. e. the null-state
$|0\rangle$  into eigenstates with \(S=-1\), i. e. the one-state
$|1\rangle$ and vice versa.

With this notation the square root of not is represented by the unitary
matrix $U[1]$:
\[ U=\frac{1}{2}\left[
     \begin{array}{cc}
       1-i & 1+i \\  1+i & 1-i
     \end{array}  \right] \ , \]
that solves:
\begin{eqnarray*}
  U^{2}|0\rangle & = & |1\rangle  \\
  U^{2}|1\rangle & = & |0\rangle
\end{eqnarray*}

In such way this supersymmetric double degeneracy represents two level
quantum system and we can see the following:

\begin{eqnarray*}
 Q & = & Q^{+}+Q^{-}= \left(
     \begin{array}{cc}
       0 & a^{+} \\  a^{-} & 0
     \end{array}  \right) \ , \\
  \tau & = & \sigma_{3} = \left(
     \begin{array}{cc}
       1 & 0  \\  0 & -1
     \end{array}  \right) \ , \\
 \{Q,\tau\} & = & 0 \ .
\end{eqnarray*}

It implies that the operator supercharge $Q$ really transforms the state
\(|0\rangle,\ |1\rangle\) as the operator square root of not quantum
algorithm operator $U^{2}$.
In such a way supersymmetric ``square root'' corresponds the
``square root of not''

\section{The information quarks}

The quantum information-theoretical formalism defined above can be generalized
to multipartie system.Consider for example a triepartie entangled quantum system u, d, s.
The chain rule for quantum entropies is
\begin{equation}
S_Q(uds)=S_Q(u)+S_Q(d) +S_Q(s|ud).
\label{3.3}
\end{equation}

The von Neumann conditional mutual entropy

\begin{equation}
S_Q(u:d|s)=S_Q(u|s) + S_Q(d|s) - S_Q(ud|s) = S_Q(us) + S_Q(ds) - S_Q(s) - S_Q(uds).
\label{3.3}
\end{equation}
which reflects the quantum mutual entropy between u and d when s is known.

For ternary mutual quantum entropy $S_Q(u:d:s)$, i.e., that piece of
the mutual entropy between u and d that is also shared by s.
It has no classical counterpart - for any entangled tripartite system uds
in a pure state, the ternary mutual entropy vanish, i.e.

\begin{equation}
S_Q(u:d:s)=S_Q(u) + S_Q(d) + S_Q(s) - S_Q(ud) - S_Q(us) - S_Q(ds) +S_Q(uds)= 0.
\label{3.3}
\end{equation}

This results from the fact that $S_Q(uds)=0$ implies$S_Q(ud)=S_Q(s)$ ,
$S_Q(us)=S_Q(d)$ , $S_Q(ds)=S_Q(u)$ , as a consequence of the Schmidt
decomposition of the state uds.

\section{Supersymmetric qubit fields}

Consider commuting or anticommuting fields which can be coupled via supersymmetry. Let an field of
identically-constituted guantum physical systems on spacetime. Following \cite{8}
each event is associated with one such system and each of systems has the same
algebra of observables and obeys the same dynamical law, the field is continuous
in the sense that the fields corresponding physical quantities of each system are
continuous and also differentiable.

We shall define the superspace
$(x,\theta,\bar{\theta})$ and let $\Phi_(x,\theta,\bar{\theta})$ be the superqubit field at event $(x,\theta,\bar{\theta})$,
which fullfils:
\begin{equation}
\Phi_j \Phi_k  =  \delta_jk I + i {\epsilon_jk}^l \Phi_l
                                   \label{3.3}
\end{equation}

and so for the first  component in the expansion we obtain exactly the scalar
qubit field as D.Deutsch in \cite{8}.

The chiral superqubit field on the superspace $(x,\theta)$:
\[ \Phi(x,\theta)=\varphi(x)+i\,\theta^{\alpha}\psi_{\alpha}(x)
    +\frac{i}{2}\,\theta^{\alpha}\theta_{\alpha}\,F(x)\]
the supermultiplet $\{\varphi,\psi,F\}$ of ordinary commuting and anticommuting
qubit fields (at this moment we do not write the qubit field indeces j,k,l... without
lost of generallity)

We can generalize the super $O(3)$ $\sigma$ model:

\begin{eqnarray}
 \lefteqn{\frac{1}{2}\int d^{2}x\,d^{2}\theta\,\varepsilon^{\alpha\beta}\,
   (D_{\alpha}\Phi^{a}\cdot D_{\beta}\Phi^{a})
     \ , }                           \label{3.1}  \\
\lefteqn{\mbox{with the condition:}\qquad
  \Phi^{a}(x,\theta)\cdot\Phi^{a}(x,\theta)=1,\qquad a=1,2,3.}
                                     \label{3.2}
\end{eqnarray}

For qubit fields we obtain the following Lagrangian:
\begin{equation}
 L=\frac{1}{2}\,[\partial_{\mu}\varphi^{a}\cdot\partial_{\mu}\varphi^{a}
  +i\,\psi^{a}\cdot\gamma^{\mu}\partial_{\mu}\psi^{a}-F^{a}\cdot F^{a}]
                                   \label{3.3}
\end{equation}
and the condition gives:
\begin{eqnarray}
  \varphi^{a}\cdot\varphi^{a} & = & 1 \ ,   \nonumber      \\
  \varphi^{a}\cdot\psi^{a} & = & 0 \ ,   \label{3.4}       \\
  \varphi^{a}\cdot F^{a} & = & \frac{i}{2}\,\psi^{a}\cdot\psi^{a} \ .
                                    \nonumber
\end{eqnarray}

As usual:
\begin{equation}
 L=\frac{1}{2}\,\partial_{\mu}\varphi^{a}\cdot\partial_{\mu}\varphi^{a}
  +\frac{\alpha}{2}\,(\varphi^{a}\cdot\varphi^{a}-1)
   +\frac{i}{2}\,\psi^{a}\cdot\gamma^{\mu}\partial_{\mu}\psi^{a}
   +\frac{1}{8}\,(\psi^{a}\cdot\psi^{a})^{2} \ .
                                   \label{3.5}
\end{equation}

The equation of motion follows:
\begin{eqnarray}
 \Box\varphi^{b} + (\partial_{\mu}\varphi^{a}\cdot\partial_{\mu}\varphi^{a})
  \,\varphi^{b} & = & 0 \ ,          \label{3.6a}  \\
  i\gamma\cdot\partial\psi^{b}
   +\frac{1}{2}\,(\psi^{a}\cdot\psi^{a})\,\psi^{b} & = & 0  \ .
                                     \label{3.6b}
\end{eqnarray}

We can see from (\ref{3.6a}) that anticommuting qubits fields which are given via
supersymmetric extension do not change the equation for commuting qubit.

The supercurrent has the form:
\begin{equation}
 J_{\mu}=\gamma^{\nu}\partial_{\nu}\varphi^{a}\cdot\gamma^{\mu}\psi^{a}
                                   \label{3.7}
\end{equation}
and:
\begin{eqnarray*}
 \theta_{\mu\nu} & = &
      \partial_{\mu}\varphi^{a}\cdot\partial_{\nu}\varphi^{a}
   -\frac{1}{2}\,g_{\mu\nu}
        \partial_{\lambda}\varphi^{a}\cdot\partial_{\lambda}\varphi^{a}+ \\
   & & \mbox{}+\frac{i}{4}\,[\psi^{a}\cdot(\gamma_{\mu}\partial_{\nu}
        -\gamma_{\nu}\partial_{\mu})\,\psi^{a}
         -g_{\mu\nu}\psi^{a}\cdot\gamma^{\rho}\partial_{\rho}\psi^{a}]  \ .
\end{eqnarray*}

The super action (\ref{3.1}) and (\ref{3.2}) are invariant under the supertransformation:
\begin{eqnarray}
 \delta\varphi^{a} & = & i\varepsilon\psi^{a} \ ,   \nonumber   \\
 \delta\psi^{a} & = & (\gamma\cdot\partial\varphi^{a}+F)\,\varepsilon \ ,
                                \label{3.8}  \\
 \delta F^{a} & = & i\varepsilon\gamma^{\mu}\partial_{\mu}\psi^{a} \ .
              \nonumber
\end{eqnarray}

It is known that (\ref{3.1}) is invariant under the extended supersymmetry
with the internal symmetry $O(2)$ in the Grassmann variable.

From this complex supersymmetry the current follows:
\begin{eqnarray}
 \tilde{J}_{\mu} & = &
       \varepsilon^{abc}\varphi^{a}\partial\varphi^{b}\gamma_{\mu}\psi^{c}
       \ , \qquad a,b,c=1,\ldots,3,    \label{3.9}  \\
  V_{\mu} & = &
       \varepsilon^{abc}\varphi^{a}\psi^{b}\gamma_{\mu}\psi^{c}
                                       \label{3.10}
\end{eqnarray}
and an axial current $A_{\nu}=\varepsilon_{\mu\nu}V^{\mu}$.

We show that the actio (\ref{3.1}) is invariant under the extended supersymmetry
\(N=2\) .

We shall write the superaction (\ref{3.1}) for
super CP$^{1}$ model.

We shall define the complex qubit superfield
superpole $\Phi_{i}(x,\theta)$, \(i=1,2\), where
$\{\varphi_{i},\psi_{i},F_i\}$  are complex qubit fields and $\theta$ is
the real two component spinor.

Transformation $\Phi^{a}=\Phi_{i}\sigma_{ik}^{a}\Phi_{k}$  gives:
\begin{equation}
 S=\frac{1}{4}\int d^{2}x\,d^{2}\theta\,\overline{\nabla\Phi}_{i}\cdot
 \nabla\Phi_{i} \ ,                        \label{3.12}
\end{equation}
where \(\nabla=D-A\) is  $U(1)$ gauge invariant
covariant derivative.

Where $A$ denotes a Fermi qubit superfield, which transforms as
Abelian gauge field under $U(1)$ gauge transformation.

The condition $\Phi^{a}\cdot\Phi^{a}=1$ has the form
 $\bar{\Phi}_{i}\cdot\Phi_{i}=1$ and $A$ can be eliminated as usual:
\begin{equation}
  A = \bar{\Phi}_{i}\cdot D\Phi_{i} \ .                   \label{3.13}
\end{equation}

So the super CP$^1$ action has the form of the super
$O(3)$ $\sigma$ model action.
We can se it more directly:
\[ C(x,\theta,\bar{\theta})\equiv
    C(x,\theta_{1}+i\theta_{2},\theta_{1}-i\theta_{2}) \ . \]

Super transformation in $(x,\theta,\bar{\theta})$
has the form:
\[ \delta x=-\frac{i}{2}\,[\varepsilon\gamma\bar{\theta}
   +\bar{\varepsilon}\gamma\theta] \ , \ \delta\theta=\varepsilon\ ,
  \ \delta\bar{\theta}=\bar{\varepsilon},  \]
and acting on  $C(x,\theta,\bar{\theta})$:
\begin{equation}
  \delta C=[\varepsilon Q+\bar{\varepsilon}\bar{Q}]\,C \ ,  \label{3.14}
\end{equation}
where $Q$, $\bar{Q}$ are supercharges and $D$, $\bar{D}$ supercovariant
derivatives.
Here the ``self-duality''
is equivalent the superanalyticity :
\begin{equation}
  \bar{D}\,C(x,\theta,\bar{\theta})=0 \ .             \label{3.15}
\end{equation}

The condition (\ref{3.15}) playes here the role of the following constraint:
\begin{equation}
  C(x,\theta,\bar{\theta}) =
      C(x-\frac{i}{2}\,\bar{\theta}\gamma\theta,\theta) \ .    \label{3.16}
\end{equation}

Complex supersymmetry gives:
\begin{eqnarray}
 \delta\varphi_{c} & = & i\varepsilon\bar{\psi}_{c} \ ,   \nonumber   \\
 \delta\psi_{c} & = &
         \varepsilon F_{c}+\partial\varphi_{c}\bar{\varepsilon} \ ,
                                \label{3.17}  \\
 \delta F_{c} & = & i\bar{\varepsilon}\partial\psi_{c}
              \nonumber
\end{eqnarray}
.

We can see the following identification \(\varphi_{i}, \ \psi_{i}, \ F_{i}, \ i=1,2\)
with (\ref{3.12}) :
\begin{equation}
  \varphi_{C}=\varphi_{1}+i\varphi_{2} \ , \
  \psi_{C}=\psi_{1}+i\psi_{2} \ , \
  F_{C}=F_{1}-iF_{2} \ .
                                  \label{3.18}
\end{equation}

For complex conjugate superfield  $\bar{C}$
superaalyticity is valid:
\begin{equation}
  D\bar{C}=0 \ .                                  \label{3.19}
\end{equation}

AS usual in super CP$^{1}$ model we have
the vectorov' superfield $V(x,\theta,\bar{\theta})$,
which has in
Wess-Zumino gauge the form:
\begin{equation}
  V=\frac{1}{2}\,\bar{\theta}\gamma^{\mu}\theta\,v_{\mu}
  + \mbox{``koeficienty se dvˆma a v¡ce $\theta$''} \ ,
                                  \label{3.20}
\end{equation}
and  $C$, $V$ transforms under $U(1)$ gauge as follows
transformaci n sledovnˆ:
\[ \delta C=i\Lambda C \ , \qquad \delta\bar{C}=i\bar{\Lambda}\bar{C} \ ,
   \qquad \delta V=i(\Lambda-\bar{\Lambda}) \ , \ \]
where $\Lambda(x_{\mu}-\frac{1}{2}\,\bar{\theta}\gamma_{\mu}\theta,\theta)$
is chiral gauge supefield.

The complex supersymmetry and gauge invariant action has the form :
\begin{equation}
  S=\frac{1}{8}\int dx^{2}\,dx^{2}\theta\,dx^{2}\bar{\theta}\,
   (-V+C\bar{C}\,e^{V})
                                  \label{3.21}
\end{equation}
and equations of motion follow:
\begin{equation}
   C\bar{C}=e^{-V}=1-V \ ,
                                  \label{3.22}
\end{equation}
and for the first term in the $\theta$ expansion we get:
\[ \bar{\varphi}\varphi=1 \ , \qquad
   \bar{\psi} \cdot\varphi=\psi\cdot\bar{\varphi}=0 \ .       \]
So  $V$ acts as an unification force.

\section{Conclusions}

We obtained in scalar qubit field thery proposed by
D.Deutsch an interaction with anticommuting qubit field via natural supersymmetric way.
If qubit fields are realised in nature then also antiqubit fields must exist and
and then all existinq quantum field theories that have empirical corroboration
are presumably long range aproximations to an exact theory of interacting qubit fields.

The interisting aplication of this is that the existence of the  condensate
$\bar{\psi}\psi$ from the anticommuting qubits. It can be interpreted
as the information vacuum and it can be broken
near the black hole on qubit and antiqubit fermionic field and so the black hole
can radiate also an information.It is the quantum information analog of the Hawking radiation.
Also the holographic bound given in the work \cite{4} for the entropy S of any
region enclosed by surface of area A:

\begin{equation}
    S = \frac{Ak}{4{l_p}^2}
                                  \label{8.1}
\end{equation}
where $l_p$ is the Planck length and $k$ the Boltzmann constant.
The equality holds if and only if the system is static black hole so it means that
that a black hole has at least as much entropy as any other object that could
be enclosed in the same surface. The Hilbert space of the quantum fields
inside such region therefore cannot have dimension higher then ${\exp}^\frac{S}{k}$
and so the region cannot hold more than $\frac{Ac^4}{4hG\ln2}$ bits.This
information paradox can be explained via antiqubit radiation of information from
the black hole. Entropy inside black hole plus outside must increase.
At least here we show an example of theory of interacting qubit fields which
can be extended.

\end{document}